# Magnetic fingerprint of interfacial coupling between CoFe and nanoscale ferroelectric domain walls

Qintong Zhang[1], Peyton Murray[2], Lu You[3], Caihua Wan[1], Xuan Zhang[1], Wenjing Li[1], Usman Khan[1], Junling Wang[3], Kai Liu[2,*] and Xiufeng Han[1,*]

[1]Beijing National Laboratory for Condensed Matter Physics, Institute of Physics, Chinese Academy of Sciences, Beijing 100190, China

[2]Physics Department, University of California, Davis, California 95616, USA

[3]School of Materials Science and Engineering, Nanyang Technological University, 639798, Singapore

**KEYWORDS**: First-Order Reversal Curve, BiFeO$_3$, Multiferroics, Exchange Bias

**ABSTRACT**: Magnetoelectric coupling in ferromagnet/multiferroic systems is often manifested in the exchange bias effect, which may have combined contributions from multiple sources, such as domain walls, chemical defects or strain. In this study we magnetically "fingerprint" the coupling behavior of CoFe grown on epitaxial BiFeO$_3$ (BFO) thin films by magnetometry and first-order-reversal-curves (FORC). The contribution to exchange bias from 71º, 109º and charged ferroelectric domain walls (DWs) was elucidated by the FORC distribution. CoFe samples grown on BFO with 71º DWs only exhibit an enhancement of the coercivity, but little exchange bias. Samples grown on BFO with 109º DWs and mosaic DWs exhibit a much larger exchange bias, with the main enhancement attributed to 109° and charged DWs. Based on the Malozemoff random field model, a varying-anisotropy model is proposed to account for the exchange bias enhancement. This work sheds light on the relationship between the exchange bias effect of the CoFe/BFO heterointerface and the ferroelectric DWs, and provides a path for multiferroic device analysis and design.

As a promising candidate for room temperature magnetoelectric devices, $BiFeO_3$ (BFO) has been extensively studied in the past decade.[1-3] Much efforts have been devoted to studying magnetic properties of BFO epitaxial films[4-6] or single crystals[7, 8] and their magnetoelectric coupling with adjacent ferromagnetic (FM) materials.[4, 9-14] These studies have been motivated by the possibility of controlling the magnetism by electrically tuning the magnitude of the exchange bias[15, 16] or tailoring the magnetic anisotropy of the FM layer.[17-19] The exchange bias effect[20] has been prominently featured as manifestation of the magnetoelectric coupling. Recently, Martin *et al.* have demonstrated that the magnitude of the exchange bias in CoFe/BFO heterostructures strongly depends on the presence of 109° ferroelectric (FE) domain walls (DWs).[11] On the other hand, Sando *et al.* have shown that the exchange bias can be tuned by epitaxial strain using Mössbauer and Raman spectroscopies, combined with Landau-Ginzburg theory and effective Hamiltonian calculations.[4] These results suggest that exchange bias in FM/BFO heterostructure is likely a manifestation of the combined contributions from DWs,[11, 21] chemical defects,[22] and epitaxial strain,[4] which are challenging to distinguish.

Due to the rhombohedral symmetry of the BFO, there are three types of ferroelectric DWs, namely 71°, 109° and 180°, depending on the relative angles of polarization vectors between adjacent domains. Unlike ferromagnetic DWs which usually adopt Bloch or Néel-type configuration of the magnetic moments, ferroelectric DWs may have mixed characters due to the strong coupling between the polarization and the lattice.[23] Specifically, perovskite oxides such as BFO exhibit complex patterns of oxygen octahedral rotation at different types of DWs, rendering their fascinating and diverse properties.[24-27] Owing to the structural and electrostatic discontinuities at the DWs, charged defects with low formation energies preferably reside therein, resulting in both intrinsic and extrinsic contributions to magnetoelectric coupling.[28-31]

Conventional studies of exchange bias usually rely on major loop measurements which only capture the ensemble-averaged exchange field ($H_E$) and coercivity ($H_C$), making it difficult to distinguish contributions from various sources or distributions of exchange bias and local coercivity. In this study, the first-order reversal curve (FORC) method was employed to investigate interfacial coupling behaviors in CoFe/BFO heterostructures mediated by ferroelectric DWs. The FORC distributions map out the variations in exchange bias and coercivity, and establish the contributions from various ferroelectric DWs. These findings provide insights for multiferroic device concepts and applications.

(001)-oriented epitaxial $BiFeO_3$ films, 50 nm in thickness, with different types of domain walls were fabricated by pulsed laser deposition (PLD) on (011)-oriented $DyScO_3$ (DSO) single crystal substrates held at 650 °C to 700 °C using a stoichiometric $BiFeO_3$ target. BFO films were grown under 13 Pa oxygen pressure and *in-situ* cooled down to room temperature in an oxygen atmosphere at $10^4$ Pa for 2 hours. By increasing the substrate temperature during growth, the domain structure of BFO can be controlled from exhibiting 71° to 109° DWs, even to a mosaic structure.[32] A 3-nm-thick CoFe and a 5-nm-thick Pt capping layer were then grown on BFO films by PLD in a magnetic field of 100 Oe. Surface topography and ferroelectric domain images were studied using piezoelectric force microscopy (PFM) based on an atomic force microscope system (Asylum Research MFP-3D).

Magnetometry measurements were performed using vibrating sample magnetometry (VSM), in conjunction with the FORC method.[33-36] FORC analysis involves measurements of many partial hysteresis curves, or FORC's, each starting at a progressively more negative reversal field ($H_r$) after positive saturation, and measuring the magnetization, $M$, as the applied field ($H$) is increased back to saturation. The FORC distribution $\rho$ is then extracted using a mixed partial derivative,

$$\rho = -\frac{1}{2}\frac{\partial^2 M\left(H, H_r\right)}{\partial H \partial H_r} \qquad (1)$$

which eliminates purely reversible component of the magnetization switching. FORC analysis is useful not only in making direct measurements of the exchange bias,[37] but also in identifying the dominant interactions in a system,[35] or distinguishing the presence of multiple magnetic phases.[38] Alternatively, the FORC distribution can be plotted in another set of coordinates: the local coercivity, $H_C = (H-H_r)/2$, and the bias field, $H_B = (H+H_r)/2$.

The topographic and ferroelectric domain images of BFO films were obtained by atomic force microscopy [Figs. 1(a), (d), (g)] and piezoelectric force microscopy, respectively. The rms roughness of all BFO thin films was about 0.3 nm. Combining the in-plane [Figs. 1(b), (e), (h)] and out-of-plane [Figs. 1(c), (f), (i)] PFM images, three types of ferroelectric domain patterns for BFO films grown on DSO substrates were demonstrated, respectively, following the same protocols described elsewhere.[39, 40] Sample 1 contains mostly 71$^\circ$ DWs. Also observed were discontinuities in stripes of 71$^\circ$ DWs, which appear as branch points and end points[41] and can be generally identified as charged DWs (CDWs) that have head-to-head or tail-to-tail ferroelectric polarization configurations. Examples of branch/end points are illustrated by blue/red circles in Fig. 1(b). Counting the number of branch/end points in DWs in each sample gives a quantitative estimate of the number of CDWs. The PFM images of sample 1 [Figs. 1(b), 1(c)] show a total of 86 branch/end points. In contrast, sample 2 exhibits a mixture of 71$^\circ$, 109$^\circ$, and CDWs, with 156 branch/end points. Sample 3 has a mosaic of all DW types. The small size of the domain mosaic [Figs. 1(h), 1(i)] makes it difficult to count the number of branch/end points in this sample. However, the complex domain pattern as well as the large number of "speckle" domains suggests a larger number of CDWs than in either sample 1 or 2.

To understand the influence of DWs on the exchange coupling behavior of CoFe/BFO, room-temperature major magnetic hysteresis loops were measured on CoFe grown on the aforementioned three separate BFO films, each with a different DW pattern, along with a reference sample of CoFe grown on a bare DSO substrate. The CoFe/DSO(sub.) sample shows an unbiased loop with a coercivity of 14 Oe [Fig. 2(a)]. The CoFe/BFO sample 1 exhibits not only an enhanced coercivity of 28 Oe and a small bias of 6 Oe, but also a kinked hysteresis loop, indicating the presence of a second phase [Fig. 2(b)]. The magnitude of the exchange bias further increases to 24 Oe, without the kink in the loop shape for sample 2 [Fig. 2(c)], and to 35 Oe in sample 3 [Fig. 2(d)].

FORC distributions of the three CoFe/BFO samples are shown in Fig. 3. The ferroelectric DWs have remarkably different impacts on the CoFe magnetization reversal. In the case of sample 1, the FORC distribution shows two localized features [Fig. 3(a)]: a primary peak centered at ($H_C$=32 Oe, $H_B$=6 Oe) and a smaller secondary peak at ($H_B$=18 Oe, $H_B$=24 Oe). By selectively integrating over individual features in the FORC distribution, it is possible to extract the relative fractions of the magnetic phases responsible for each feature.[38, 42] Using this approach, the contribution to the irreversible switching from the main peak was found to be 79%, with the smaller peak contributing 18%. According to previous studies by Martin *et al.*,[11] the magnitude of the exchange bias is primarily determined by the presence of 109° DWs, and 71° DWs to a much lesser extent. Thus the primary FORC peak in sample 1, with negligible bias, can be attributed to the coupling between the CoFe layer and 71° DWs, while the secondary peak can be attributed to the charged DWs, based on the observation of branch/end points in PFM images discussed earlier. The presence of these two FORC features accounts for the kink observed in the major loop shown in Fig. 2(b), with the coupling to 71° DWs contributing mostly to the CoFe coercivity but little exchange bias, and that to the charged DWs contributing to a much larger exchange bias but lower coercivity. The weighted average coercivity and exchange bias from these two phases are 29 Oe and 9 Oe, respectively, which agree quite well with major loop values, as shown in Table 1.

For sample 2, there is again a localized peak centered at ($H_C$=32 Oe, $H_B$=5 Oe) [Fig. 3(b)], near the same location as the main FORC peak observed in sample 1 with mostly 71° DWs, and thus can be attributed to the coupling between CoFe and 71° DWs. This phase contributes to 17% of the irreversible phase fraction. Additionally, a relatively weaker but broad FORC ridge is observed along the $H_B$ direction. This ridge is roughly divided into the two sections indicated in Fig. 3(b): a lower portion centered at ($H_C$=32 Oe, $H_B$=18 Oe), accounting for 34% of the phase fraction, and an upper portion centered at ($H_C$=27 Oe, $H_B$=29 Oe), accounting for 47%. Similar to the analysis of sample 1, this FORC ridge with larger $H_B$ could be attributed to the 109° and charged DWs. The weighted average coercivity and exchange bias from these 3 contributions are 29 and 22 Oe, respectively, again in good agreement with the major loop values (Table 1). Note that the exchange bias enhancement is correlated with an increase not only in the number of 109° DWs, but also the number of CDWs, which is evidenced by the larger number of DW branch/end points as compared to sample 1.

For sample 3, a narrow ridge along the $H_B$ direction is observed, extending continuously from 20 Oe to 45 Oe [Fig. 3(c)]. The uniform FORC ridge corresponds to a narrow local coercivity distribution and a range of bias fields, often characteristic of a demagnetizing interaction.[35] The bottom end of the ridge is located near the same position as that of the charged DWs in Fig. 3(a), while the top end extends to higher bias fields than those in samples 1 and 2. As demonstrated by PFM studies [Figs. 1(h), 1(i)], a disordered mosaic DW pattern is present, in which ferroelectric DWs are closely and randomly oriented across the BFO film. The large number of domains, combined with the "speckle" pattern, suggests the presence of a larger number of 109º and charged DWs with varying sizes; the magnetic moments in these DWs couple together, in contrast to sample 2 where the moments associated with charged and 109º DWs are decoupled. Together with variations in this disordered DW pattern, these effects result in a range of exchange bias between DWs and CoFe, leading to a continuous ridge along the $H_B$ direction.

Many experimental and theoretical studies have demonstrated that changes in local symmetry at the DWs might significantly affect the magnetic properties, resulting in a net weak ferromagnetism.[24, 43-45] The (001) surface of BFO in the G-type antiferromagnetic (AF) structure is fully compensated. Therefore, the net moments of the DWs are most likely responsible for the observed exchange bias. Contributions from these DWs can be considered in the same fashion as those from pinned uncompensated AF interfacial moments that lead to exchange bias,[46-48] using the random field model:[49]

$$H_E = \frac{2z\sqrt{AK}}{\pi^2 M_F t_F} \qquad (2)$$

where $z$ is a number of order unity depending on the shape of the AF domains, $K$ and $A$ are the anisotropy and stiffness constant of the AF layer, $M_F$ and $t_F$ are the magnetization and thickness of the FM layer, respectively. Because the spins in each ferroelectric domain are compensated, the only enhancement to the coercivity or contribution to the exchange bias comes from net moments in DWs at the interface. Here, we believe that the key to understanding this magnetic coupling behavior is the variation in the anisotropy constant in different DWs. When $K$ is very small, $H_E$ becomes negligible, indicating that the uncompensated spins of the AFM layer will switch with and drag the FM layer under an external field. This will lead to a negligible exchange bias field and an enhanced coercivity, which is the case in CoFe coupled to 71° DWs. On the other hand, when $K$ is large enough, uncompensated spins in the DWs act as a pinned layer, leading to large exchange bias,[50, 51] as is observed in samples with 109° and charged DWs.

It is helpful to compare the ensemble-averaged VSM results with the FORC distributions to distinguish the contributions from the 109°, 71° and charged DWs (shown in Table 1). For example, in the case of sample 2, the major hysteresis loop shows a marked increase in the exchange bias compared with that seen in sample 1. If the major loop is used to analyze this sample, one would mistakenly conclude that CoFe across the entire BFO interface has been pinned by DWs, including 71° DWs, due to the observed exchange bias. However, three features present in the FORC distribution of sample 2 are located at ($H_C$=32 Oe, $H_B$=5 Oe), ($H_C$=32 Oe, $H_B$=18 Oe), and ($H_C$=27 Oe, $H_B$=29 Oe), accounting for 17%, 34%, and 47% respectively. The contributions from these FORC features are significant enough to result in an average increase in the exchange bias field measured by VSM. That is, the contribution from the ferroelectric DWs on the exchange bias effect can be identified unambiguously using the FORC method.

In summary, we have mapped out the contributions from different types of DWs to the interlayer coupling in multiferroic systems of CoFe/epitaxial BiFeO₃ thin films. For the BFO with 71° DWs, the FORC distribution shows a coercivity enhancement with very little exchange bias. For the BFO sample with 109° DWs and charged DWs, significantly larger exchange bias is observed, due to the presence of 109° and charged DWs. These DWs provide uncompensated magnetic moments that pin the adjacent CoFe, and can be accounted for using the random field model. For the BFO sample with a mosaic DW pattern, the distribution of 109° and charged DWs leads to a uniform ridge along the bias field. These findings shed light on the origin of exchange bias in such multiferroic systems and highlight the correlation with FE DWs, which opens up directions for multiferroic device concepts and applications.

The project was supported by the State Key Project of Fundamental Research and 863 Plan Project of Ministry of Science and Technology (No. 2014AA032904), the MOST National Key Scientific Instrument and Equipment Development Projects (Grant No. 2011YQ120053), National Natural Science Foundation of China (Grant No. 11434014, 11404382 and 11222432), and the Strategic Priority Research Program (B) of the Chinese Academy of Sciences (CAS) [Grant No. XDB07030200]. Work at UCD was supported by the US National Science Foundation (DMR-1008791 and DMR-1543582). L.Y. and J.L.W. acknowledge support from the Ministry of Education, Singapore under project MOE2013-T2-1-052.

**Figures captions**

**Figure 1. Topographic and ferroelectric images of the epitaxial BFO films.** AFM, in-plane PFM, and out-of-plane PFM images of BFO films grown on $DyScO_3$ substrates. Sample 1 (a-c) has mostly 71° and a few charged DWs. Sample 2 (d-f) has a mixture of 71°, 109° and charged DWs. Sample 3 (g-i) exhibits mosaic DWs. For each sample, the AFM and PFM views are over the same area. All the images areas are 3 µm × 1.5 µm, and the scale bar is 500 nm. The blue and red circles illustrate branch and end points in ferroelectric domains, respectively.

**Figure 2. Hysteresis loops of CoFe grown on BFO.** Magnetic hysteresis loops of (a) Pt/CoFe(3 nm)/DSO, and Pt/CoFe(3 nm)/BFO (50nm)/DSO with the BFO exhibiting (b) mostly 71° DWs, (c) mixture of 71°, 109° and charged DWs, and (d) mosaic FE DWs, respectively.

**Figure 3. FORC distributions.** FORC distributions of 3 nm CoFe grown on 50nm epitaxial BFO films with (a) mostly 71° and few charged DWs, (b) mixture of 71°, 109° and the charged DWs, and (c) mosaic FE DWs, respectively. The two circled regions illustrate the upper and lower parts of the extended FORC ridge.

**TABLE 1.** Comparison of coercivity and exchange bias obtained by VSM and FORC, which helps to distinguish the contributions from the 109°, 71° and charged DWs.

**Figures**

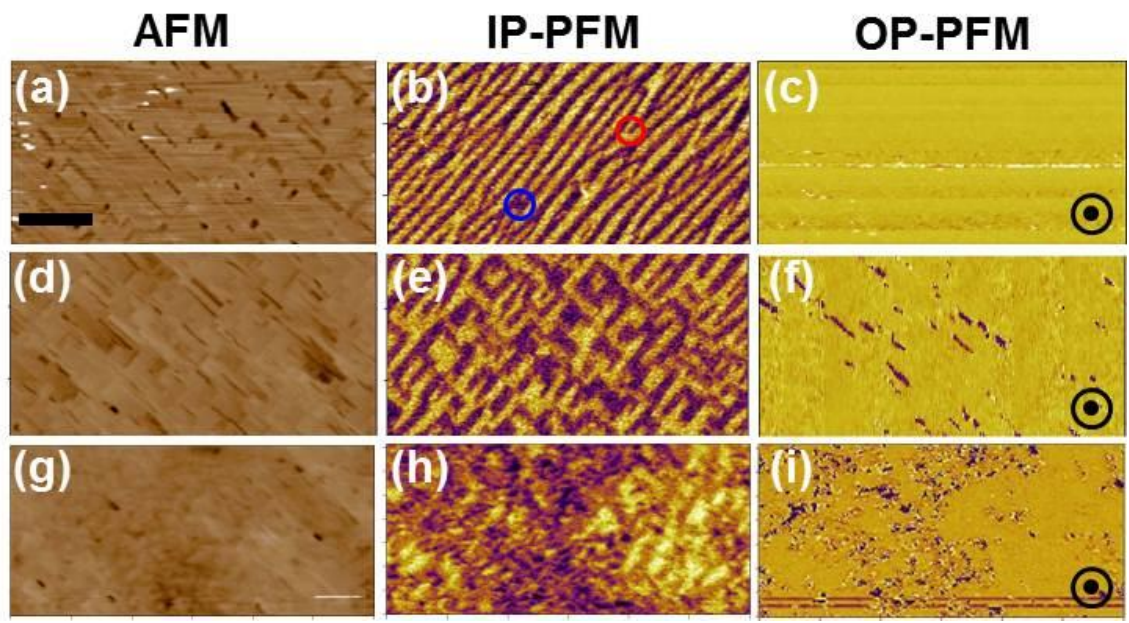

**Fig. 1**

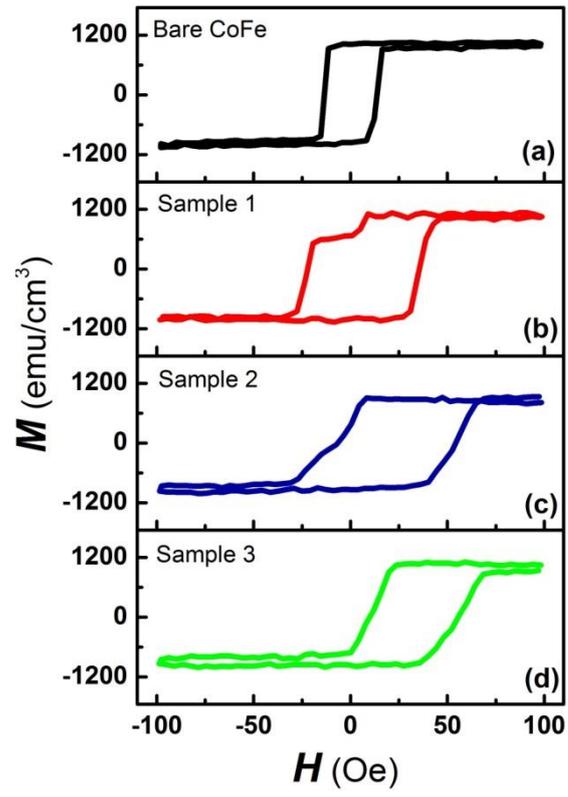

**Fig. 2**

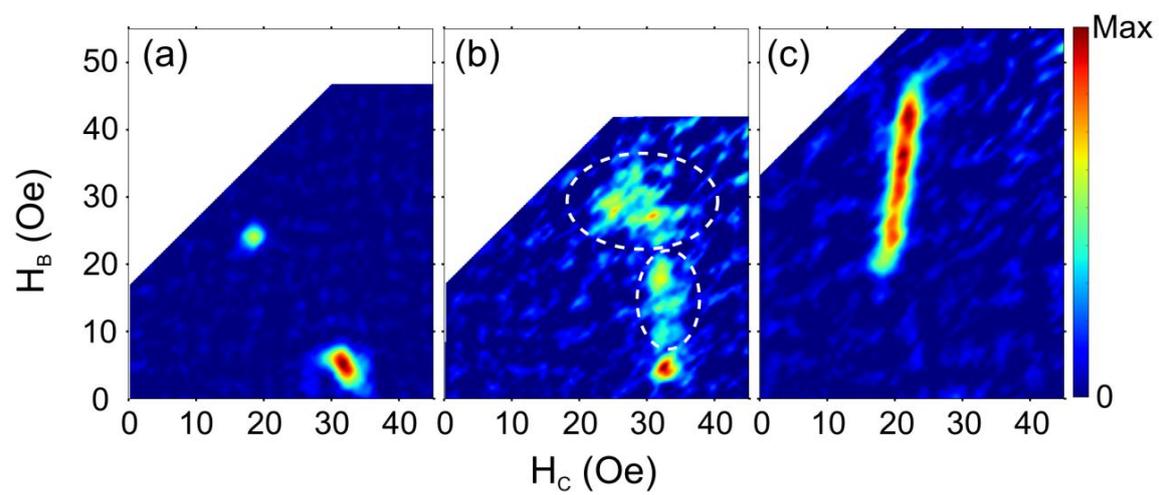

Fig. 3

**TABLE 1**

| Description | Sample 1 | | Sample 2 | | | Sample 3 |
|---|---|---|---|---|---|---|
| | 71° DWs | Charge DWs | 71° DWs | 109° DWs | Charge DWs | Mosaic |
| VSM $H_C$(Oe) | 28 | | 30 | | | 23 |
| VSM $H_E$(Oe) | 6 | | 24 | | | 35 |
| FORC Phase Fraction (%) | 79 | 18 | 17 | 34 | 47 | |
| FORC $H_C$(Oe) | 32 | 18 | 32 | 32 | 27 | 23 |
| FORC average $H_C$(Oe) | 29 | | 29 | | | |
| FORC $H_B$(Oe) | 6 | 24 | 5 | 18 | 29 | 35 |
| FORC average $H_B$(Oe) | 9 | | 21 | | | |